\documentclass[a4paper,11pt]{article}
\usepackage{pos}
\usepackage{float}
\usepackage[utf8]{inputenc}
\usepackage{natbib}
\pdfoutput=1

\title{
\begin{flushright}
\small{
WUB/21-04\\}
\vskip 0.7cm
\end{flushright}
Optimizing Distillation for charmonium and glueballs}

\author[a]{Francesco Knechtli}
\author[a]{Tomasz Korzec}
\author[b]{Mike Peardon}
\author*[a]{Juan Andrés Urrea Niño}

\affiliation[a]{Dept. of Physics, Bergische Universität Wuppertal,\\
  Gaußstrasse 20, 42119 Wuppertal, Germany}

\affiliation[b]{School of Mathematics, Trinity College,\\
Dublin 2, Ireland}

\emailAdd{knechtli@uni-wuppertal.de}
\emailAdd{korzec@uni-wuppertal.de}
\emailAdd{mjp@maths.tcd.ie}
\emailAdd{urreanino@uni-wuppertal.de}

\abstract{We study the charmonium spectrum on an ensemble with two heavy dynamical quarks with a mass at half the physical charm quark mass. Operators for different quantum numbers are used in the framework of distillation with different smearing profiles to increase the overlap with ground and excited states. The use of exact distillation, large statistics and the absence of light quarks gives robust results for the charmonium spectrum. We also present preliminary results for the glueball spectrum in this theory.}

\FullConference{%
 The 38th International Symposium on Lattice Field Theory, LATTICE2021
  26th-30th July, 2021
  Zoom/Gather@Massachusetts Institute of Technology
}

\begin{document}
\maketitle

\section{Introduction}
The quark smearing technique known as distillation has been widely used since its introduction in \cite{Peardon2009} to gain access to all-to-all smeared propagators known as \textit{perambulators} which allow to calculate improved hadronic correlation functions. The smearing is defined as a projection onto a low-dimensional space spanned by the $N_v$ lowest eigenmodes of the 3D gauge covariant Laplacian operator in a lattice of spatial volume $L^3$ and temporal extent $N_t$. These vectors are stored in a matrix $V$ of size $(4\times 3L^3 \times N_t)\times (4\times N_v \times N_t)$ whose columns are labeled $v[t]_{i,\alpha}$ corresponding to the $i$-th eigenvector of time $t$ located only in Dirac component $\alpha$ and time component $t$ and full of zeros everywhere else. The perambulators, which contain all the relevant information about quark propagation, are defined as
\begin{equation}
\tau[t_1,t_2]_{\alpha \beta}^{ij} = v[t_1]_{i,\alpha}^{\dagger}D^{-1}v[t_2]_{j,\beta}
\end{equation}
where $D$ is the Dirac operator and the \textit{elementals}, which contain all the information of the operator $\Gamma$ that defines the symmetry channel, are defined as
\begin{equation}
\Phi[t]_{\alpha, \beta}^{ij} = v[t]_{i,\alpha}^{\dagger} \Gamma v[t]_{j,\beta}.
\end{equation}
Given that the perambulators are independent of the $\Gamma$ operator, a wide variety of such operators can be used without incurring into additional inversion costs. However, the cost of calculating the perambulators, given by $4\times N_v \times N_t$ inversions of the Dirac operator, can become too high when sufficiently large lattices and high number of eigenvectors are considered. Namely, the number of eigenvectors required to keep a fixed physical eigenvalue threshold scales with the physical volume of the lattice making the study of large lattices difficult. To deal with this problem the framework of stochastic distillation was introduced in \cite{Morningstar2011}, where stochastic estimation together with dilution operators in distillation space allows to keep the number of inversions approximately fixed even as the lattice volume increases. Carefully chosen dilution schemes can be used to reduce the stochastic noise as much as possible yet the exact calculation of the perambulators is no longer performed. In this work we present an alternative improvement to the distillation method where all inversions are still performed yet the eigenvectors, specifically in the elementals, will be used in what can be considered the best way possible. 
\section{Exploiting the quark distillation profiles}
The distillation operator which acts on the quark fields at time $t$ is defined, in the most general way, as
\begin{equation}
S[t] = V[t]J[t]V[t]^{\dagger},
\end{equation}
where $V[t]$ corresponds to the $\left(4\times 3L^3\right)\times (4\times N_v)$ block of $V$ whose columns are the components at time $t$ of the vectors $v[t]_{i,\alpha}$ and $J[t]$ is a diagonal matrix acting on distillation space which is often taken as $J[t]=1$. This matrix serves to weigh the contribution that each eigenmode will have in the smeared quark field and if taken as the identity then $S[t]$ becomes an othogonal projector onto the span of the eigenmodes, defining standard distillation, but if not then $S[t]$ serves as a modulated "low eigenmode" filter. In this case the elementals become
\begin{equation}
\Phi[t] = J[t]^{\dagger}V[t]^{\dagger}\Gamma V[t]J[t]
\end{equation}
and, when defining the entries of $J[t]$ as $J[t]_{ij} = g\left( \lambda_i[t] \right) \delta_{ij}$ with $\lambda_i[t]$ the $i$-th eigenvalue of the Laplacian at time $t$ and $g(\lambda)$ the \textit{quark distillation profile}, the entries of the elemental become
\begin{equation}
\Phi[t]^{\alpha \beta}_{ij} = g(\lambda_i[t])^{*}g(\lambda_j[t]) v[t]_{i,\alpha}^{\dagger} \Gamma v[t]_{j,\beta}
\end{equation}
and $f(\lambda_i[t],\lambda_j[t]) = g(\lambda_i[t])^{*}g(\lambda_j[t])$ can be thought of as a \textit{meson distillation profile} that modulates the coupling of the eigenvectors via the $\Gamma$ operator. One can now attempt to find an optimal meson distillaton profile for different operators which might not be as simple as a product of two quark distillation profiles. This optimization is done by first defining a set of $N_{B}$ quark profiles $g_a(\lambda)$, $a=0,...,N_{B}-1$, which will serve to build different meson operators $O_{a}(t)$ where $g_a(\lambda)$ is acting accordingly on the quark and anti-quark that make up the meson. From these meson operators a correlation matrix $C(t)$ can be built as
\begin{equation}
C(t)_{ab} = \left\langle  O_{a}(t) \bar{O}_{b}(0)\right\rangle,
\end{equation}
where $\left \langle ... \right \rangle$ denotes an average over gauge configurations. For the sake of numerical stability and keeping useful operators for the calculation this matrix can be pruned \cite{Balog1999, Niedermayer2001} by projecting it onto its $N_{S}$ most significant singular vectors $u_{i}$ at a given $t_S$, resulting in
\begin{equation}
\tilde{C}(t)_{ij} = u_{i}^{\dagger} C(t) u_{j},\ t \geq t_S.
\end{equation}
At this point a generalized eigenvalue problem can be formulated with $\tilde{C}(t)$ as
\begin{equation}
\tilde{C}(t)w_k(t,t_0) = \rho_k(t,t_0) \tilde{C}(t_0) w_k(t,t_0)
\end{equation}
where from the eigenvalues $\rho_k(t,t_0)$ one can extract the effective mass of the $k$-th energy state and from the eigenvectors $w_k(t,t_0)$ one can get the coefficients that define the linear combination of the basis operators of $\tilde{C}(t)$ that yields a meson operator that overlaps the most with the $k$-th energy state \cite{Michael1983,Lscher1990,collaboration2009}. These basis operators are defined by the meson distillation profiles obtained after pruning the original $C(t)$ and therefore will be a linear combination of the original meson profiles built from simple products of the quark distillation profiles. These pruned profiles can be written explicitly as
\begin{equation}
f_{k}^{(p)}(\lambda_i, \lambda_j) = \sum_{a=0}^{N_{B}-1} u_{k}^{(a)} g_{a}(\lambda_i)^{*}g_{a}(\lambda_j)
\end{equation}
where $u_k^{(a)}$ denotes the $a$-th entry of the singular vector $u_k$ and the $(p)$ simply labels the profile as being a pruned one. The profiles corresponding to the optimal operator for energy state $e=0,...,N_{S}-1$ can be built from these pruned profiles as
\begin{equation}
\tilde{f}_{e} (\lambda_i, \lambda_j) = \sum_{k=0}^{N_{S}-1} w_{e}^{(k)} f_{k}^{(p)}(\lambda_i, \lambda_j)
\end{equation}
and inserted into what can be considered an optimal meson elemental for energy state $e$ as
\begin{equation}
\Phi_{e}[t]_{ij}^{\alpha \beta} = \tilde{f}_{e}(\lambda_i, \lambda_j) v[t]_{i,\alpha}^{\dagger} \Gamma v[t]_{j,\beta}.
\end{equation}

\section{Charmonium spectrum results}
\begin{figure}[H]
\centering
\includegraphics[width=0.6\textwidth]{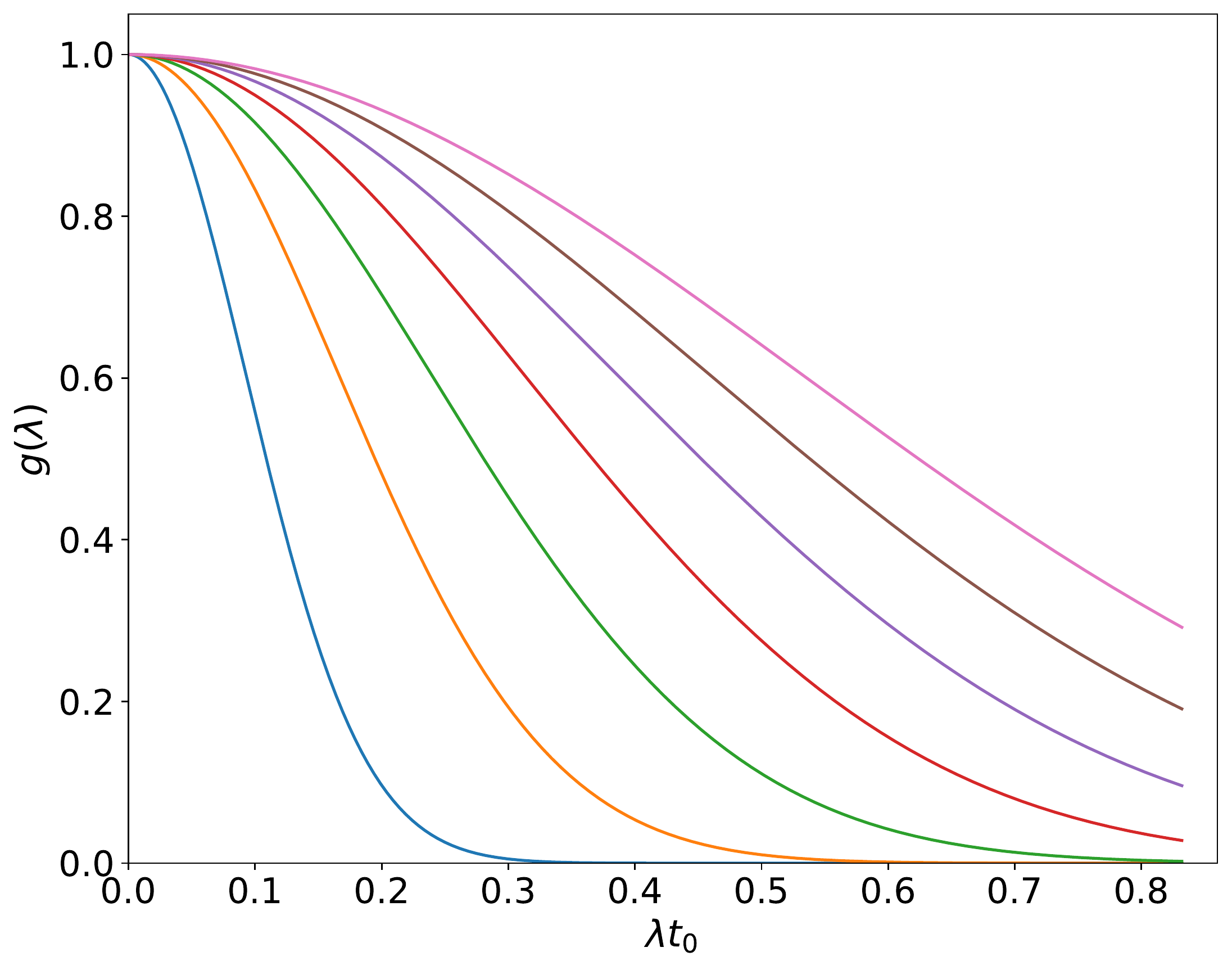}
\caption{Quark distillation profiles used in this study.}
\label{Fig:Quark_Profiles}
\end{figure}

\begin{figure}[H]
\centering
\includegraphics[width=0.6\textwidth]{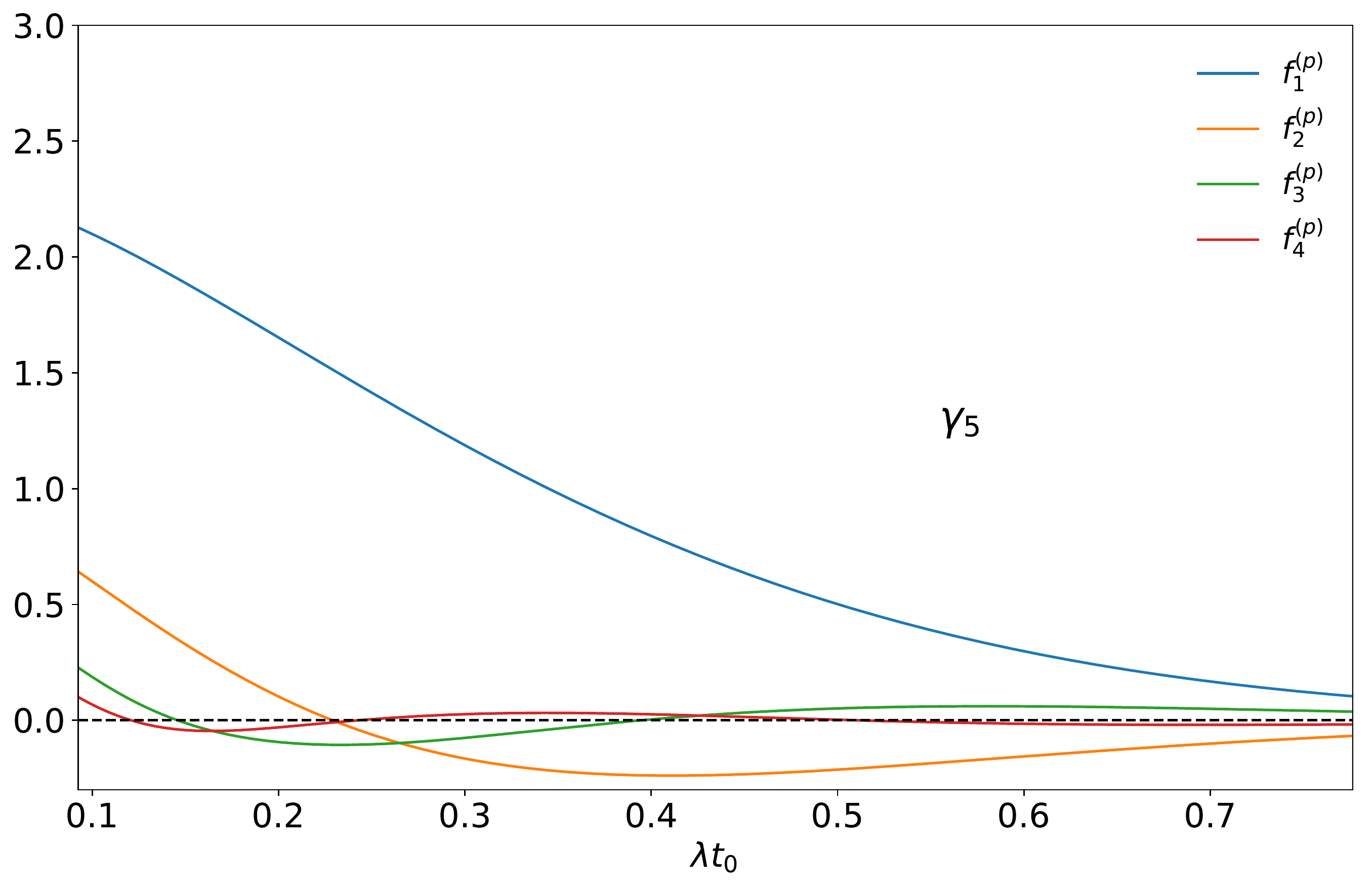}
\caption{Meson distillation profiles that make up the basis for the pruned correlation matrix $\tilde{C}(t)$ using $\Gamma = \gamma_5$.}
\label{Fig:G5_PrunedBasis}
\includegraphics[width=0.6\textwidth]{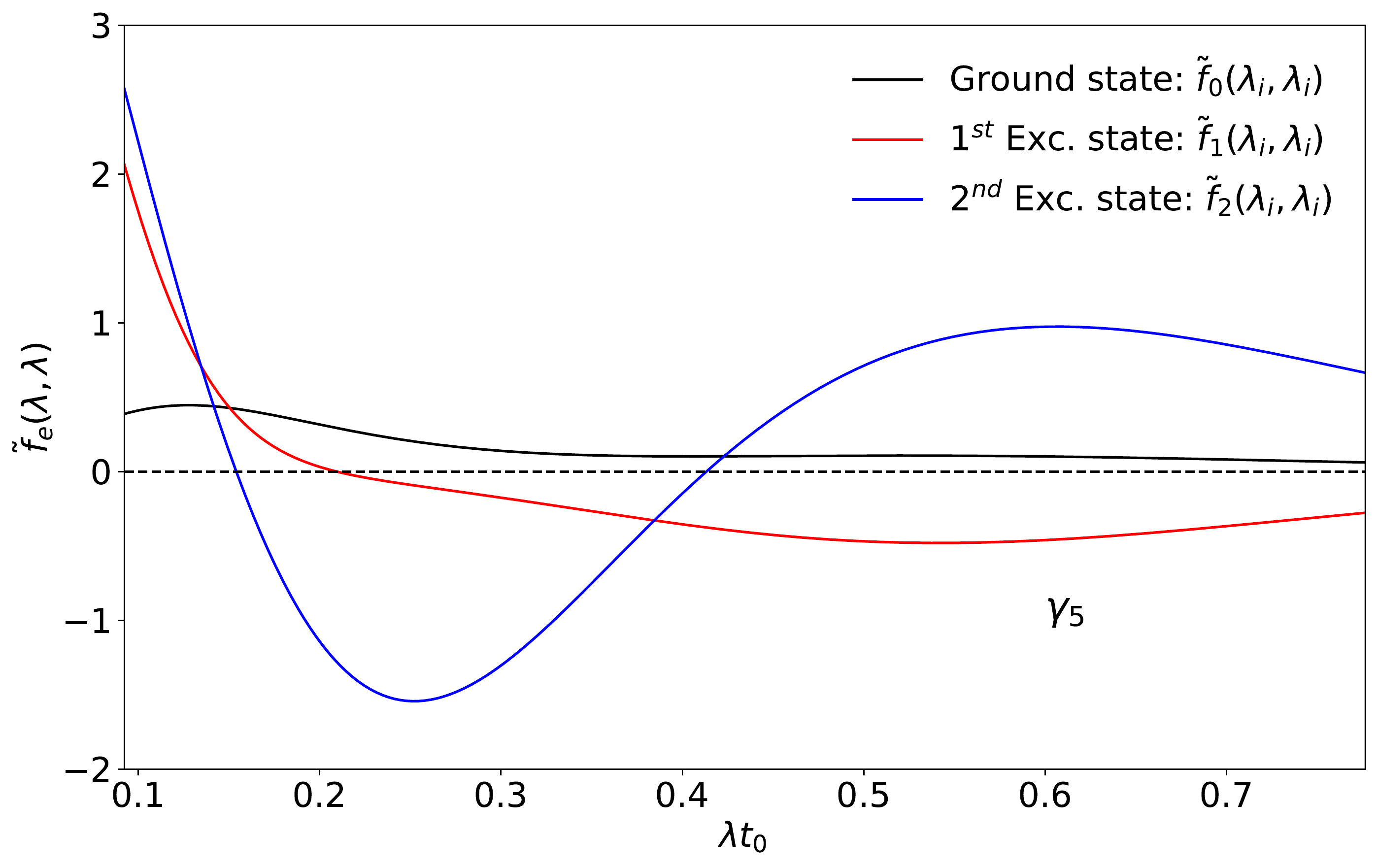}
\caption{Optimal meson distillation profiles for the three lowest states using $\Gamma = \gamma_5$.}
\label{Fig:G5_OptimalProfiles}
\end{figure}

\begin{figure}[H]
\centering
\includegraphics[width=0.6\textwidth]{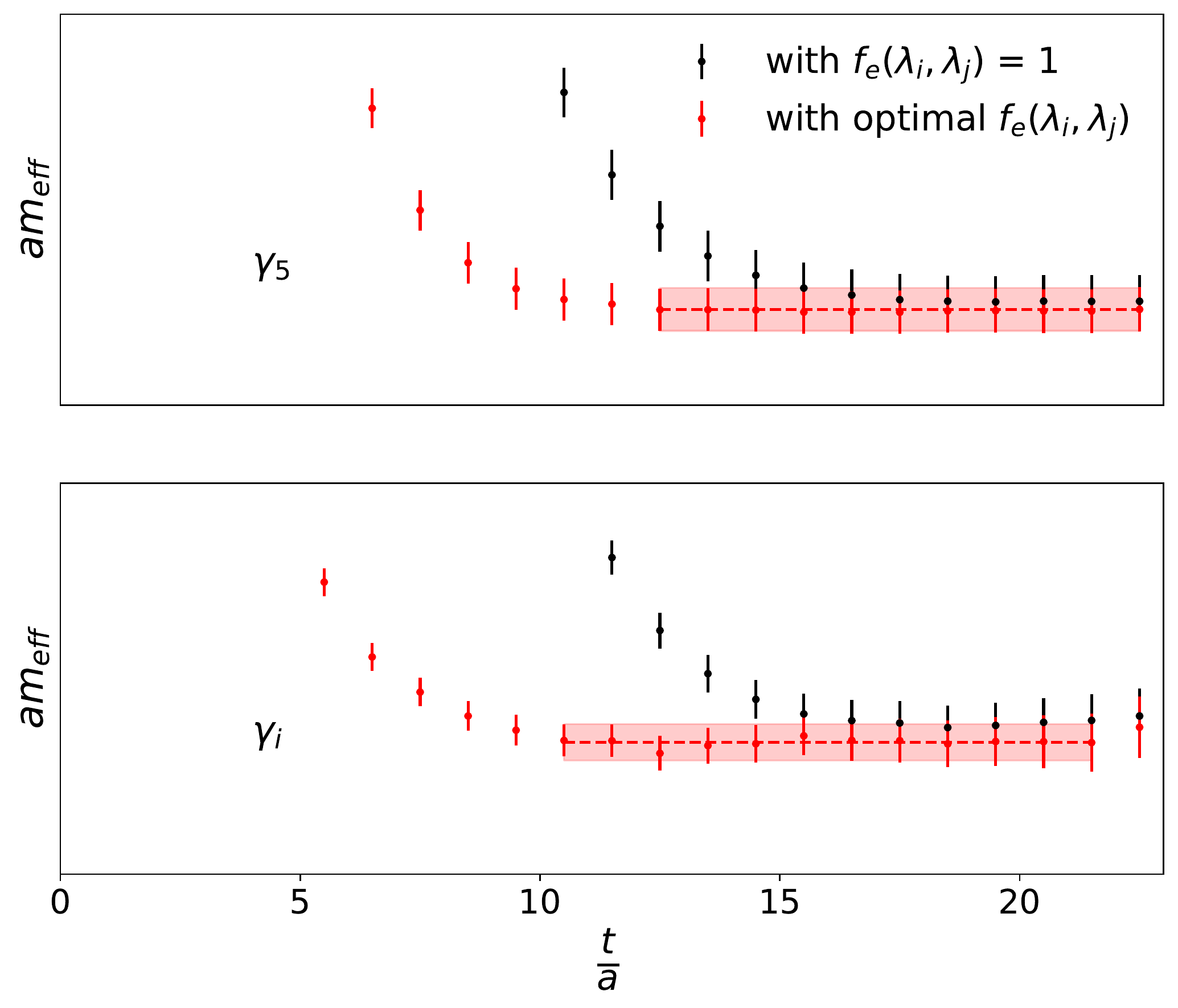}
\caption{Ground state effective masses of $\Gamma = \gamma_5,\gamma_i$ using standard distillation and the optimal meson distillation profile.}
\label{Fig:Mass_Comparison}
\end{figure}

\begin{figure}[H]
\centering
\includegraphics[width=0.6\textwidth]{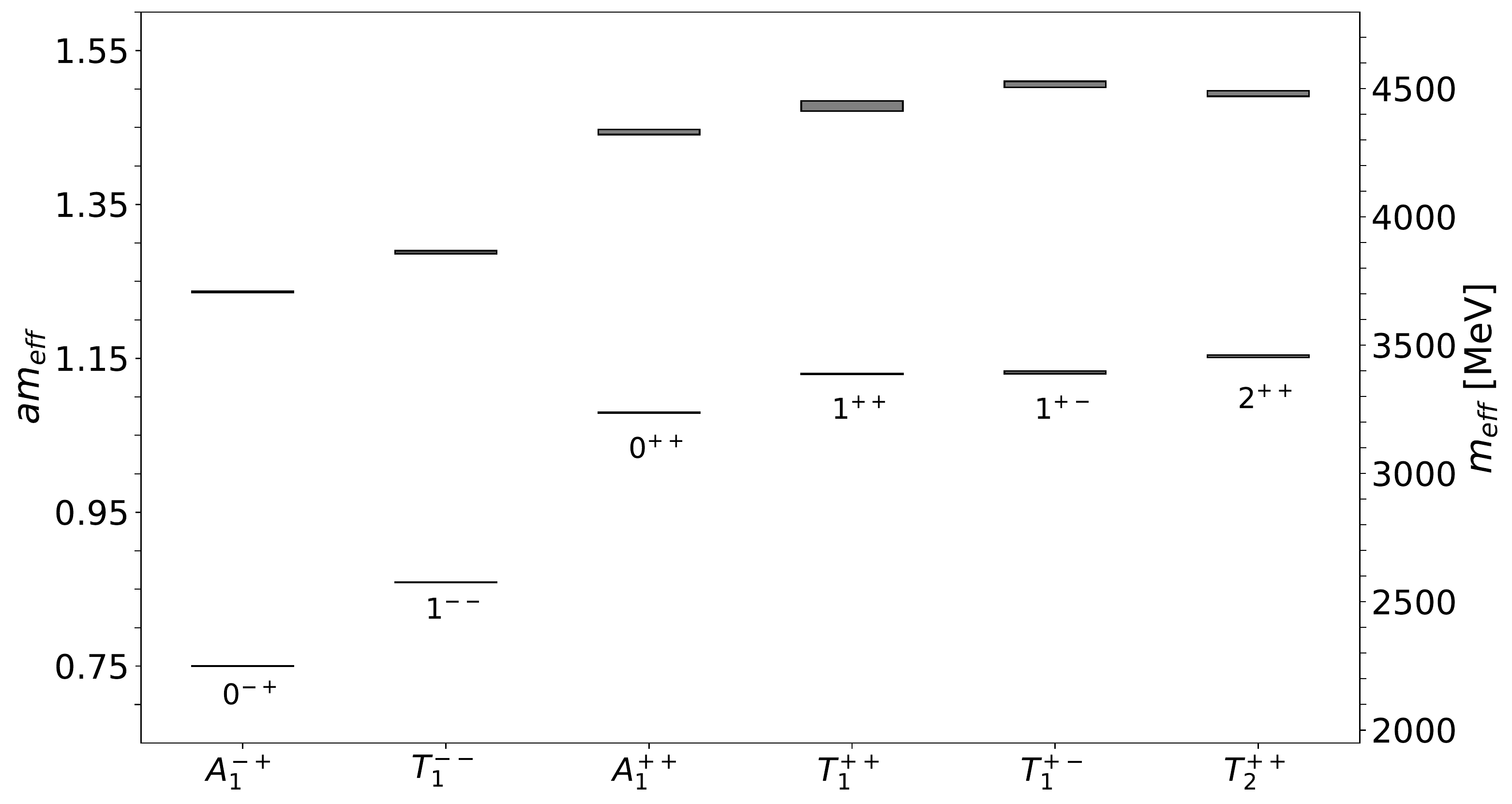}
\caption{Lowest two states for different irreps of interest.}
\label{Fig:Charm_Spectrum}
\end{figure}

\begin{figure}[H]
\centering
\includegraphics[width=0.6\textwidth]{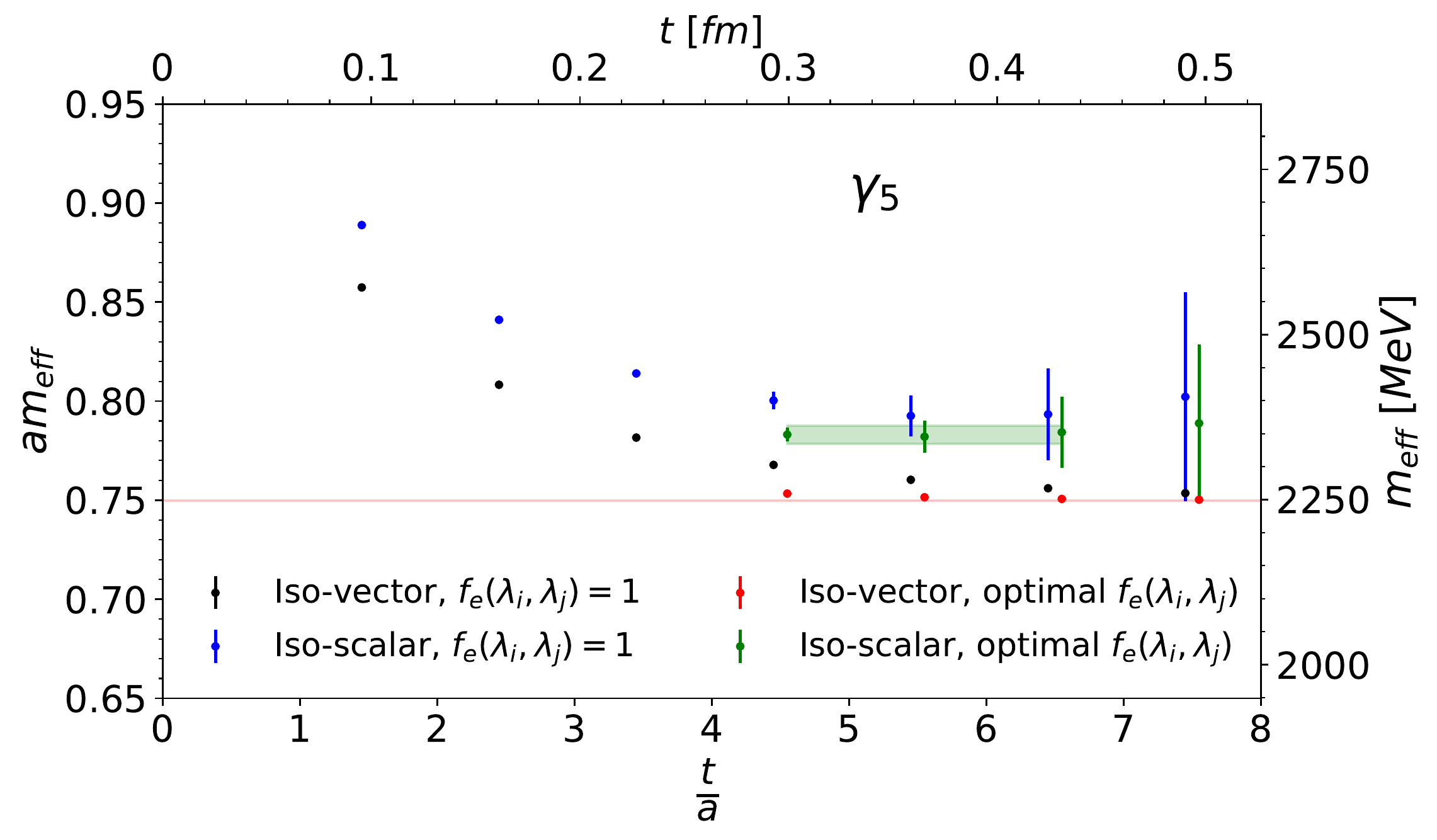}
\caption{Ground state mass for iso-scalar and iso-vector channels for $\Gamma = \gamma_5$ using standard distillation (blue and black points) and optimal meson distillation profiles (green and red points). The optimal profile in both iso-scalar and iso-vector cases corresponds to the one obtained from a GEVP using only the iso-vector data.}
\label{Fig:ScalarVsVector}
\end{figure}

The construction of an optimal meson distillation profile is tested in an ensemble with two degenerate heavy dynamical quarks with mass equal to half the mass of the physical charm quark. The lattice size is $48\times 24^3$ with a lattice spacing $a \approx 0.066$ fm and $\frac{t_0}{a^2} \approx 1.849$, where $t_0$ is the flow scale \cite{Flow}. A total of $N_v=200$ eigenvectors of the 3D gauge covariant Laplacian operator are used per time slice. The quark distillation profiles used are given by $N_{B} = 7$ gaussians of the form
\begin{equation}
g_{k}(\lambda) = e^{-\dfrac{\lambda^{2}}{2\sigma_{k}^{2}}},
\end{equation} 
where $k=0,...,N_{B}-1$ and the different widths $\sigma_k$ are given by
\begin{equation}
\sigma t_0 = [0.092,0.165,0.238,0.311,0.384,0.457,0.529].
\end{equation}
These choices for the widths can be graphically justified by looking at Fig. \ref{Fig:Quark_Profiles}, where it can be seen that the chosen widths allow for a wide range of decays over the 200 eigenvalues considered in this study. Other choices of $g_{k}(\lambda)$ were tested, such as $g_{k}(\lambda) = \lambda^{k}$, yet the gaussian basis yielded the most numerically stable results.
Once the quark distillation profiles are chosen the optimal meson profiles for a given $\Gamma$ operator can be found. For the case of $\Gamma = \gamma_5$ Fig. \ref{Fig:G5_PrunedBasis} shows the pruned meson profiles for $N_{S} = 4$ and Fig. \ref{Fig:G5_OptimalProfiles} shows the resulting optimal profiles for the lowest three states when the GEVP is set with $t_0 = t_S = 3$. Since $\gamma_5$ is a local operator the elemental is diagonal in distillation space and only $\tilde{f}_{e}(\lambda_i, \lambda_i)$ matters, which can be plotted solely as a function of $\lambda_i$. First, it is clear that the optimal meson profiles are not a constant, contrary to what is enforced in standard distillation, and they are different for different energy states, which is also not present in standard distillation. Second, especially for the ground state, low eigenmodes have larger contributions than the high ones, indicating that a suppression of high eigenmodes beyond the projection that standard distillation performs is convenient. Fig. \ref{Fig:Mass_Comparison} shows a comparison of the ground state effective mass for $\Gamma = \gamma_5, \gamma_i$ using standard distillation and with the optimal meson distillation profile, displaying the suppressed excited state contamination via an earlier mass plateau and the success of this improvement.
\vspace{2pt}
Fig. \ref{Fig:Charm_Spectrum} displays the effective masses, extracted from the corresponding plateaus, for the two lowest states of different irreps and quantum numbers of interest. All of these results, except for the case of $T_2$, were obtained with local operators and using a total of 4080 gauge configurations. The $T_2$ operator corresponds to $|\epsilon_{ijk}|\gamma_j \nabla_k$, with $\nabla_k$ being the symmetric covariant lattice derivative in spatial direction $k$ \cite{Dudek2008}, and is analyzed over a subset of 1500 gauge configurations. The use of the optimal meson distillation profiles allows to obtain clear results in all of the analyzed channels.
A quantity of special interest is the mass difference between the iso-scalar and iso-vector channels for the pseudoscalar case. In this study this quantity is extracted in the following manner. First, the optimal meson distillation profile for the iso-vector is calculated via the pruning and GEVP previously described. Second, this same profile is used for the calculation of the iso-scalar correlation. Finally, the corresponding effective masses are extracted and, after determining an early time plateau for the iso-scalar case, the mass difference is calculated. The optimal meson distillation profile for the iso-scalar is not calculated because that correlation data are considerably noisier than the iso-vector case, which has a significant negative impact on the GEVP and the results it yields. There is no reason to assume this optimal meson distillation profile should be identical to the iso-vector one yet one may suspect that it is closer to the the iso-vector one than to the constant that standard distillation assumes. Fig. \ref{Fig:ScalarVsVector} shows the ground state effective masses for both iso-scalar and iso-vector with $\Gamma = \gamma_5$ using both standard distillation and the optimal meson profiles. A reduction of excited state contamination in both cases is seen again, confirming the suspicion that using the profile of the iso-vector for the iso-scalar is better than using a constant. The iso-scalar effective masses are considerably noisier than their iso-vector counterparts so a preliminary plateau is taken at early times keeping in mind that excited state contamination is still present. The value of the mass difference in this work is found to be $99 \pm 15$ MeV, which is non-negligible, and the use of larger statistics could lead to a more precise determination. An indirect measurement of this same quantity, although in a $N_f = 2+1+1$ QCD + QED setup with different quark masses (physical charm), yielded a value of $7.3\pm 1.2$ MeV \cite{Hatton2020}. The difference in quark masses can account for the difference between this mass splitting and the one presented in this work.

\section{Glueball results}
\begin{figure}[H]
\centering
\includegraphics[width=0.6\textwidth]{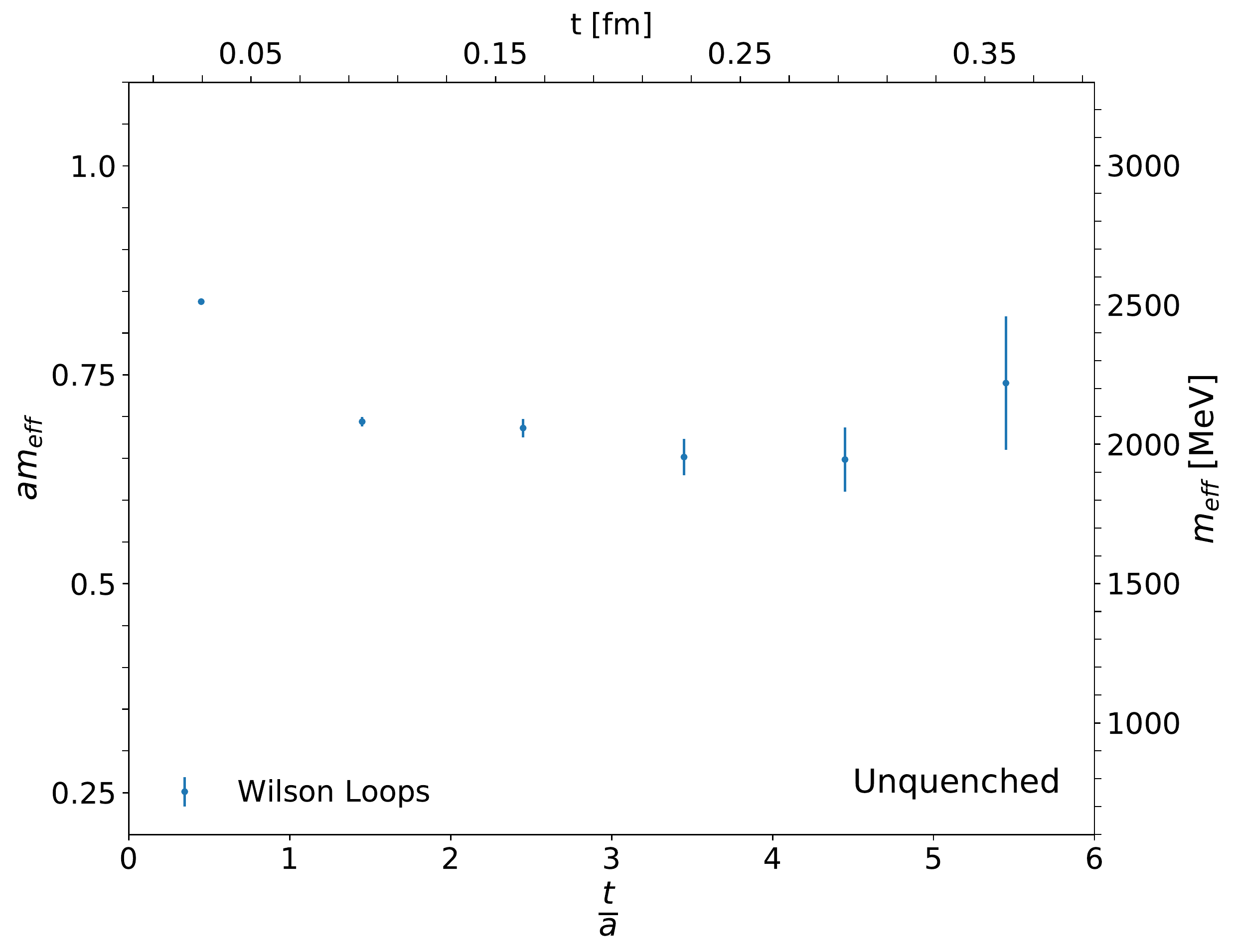}
\caption{Ground state effective mass of the scalar glueball made up of Wilson loops in the setup with 2 dynamical charm quarks.}
\label{Fig:UnquenchedScalarGlueball}
\end{figure}

\begin{figure}[H]
\centering
\includegraphics[width=0.6\textwidth]{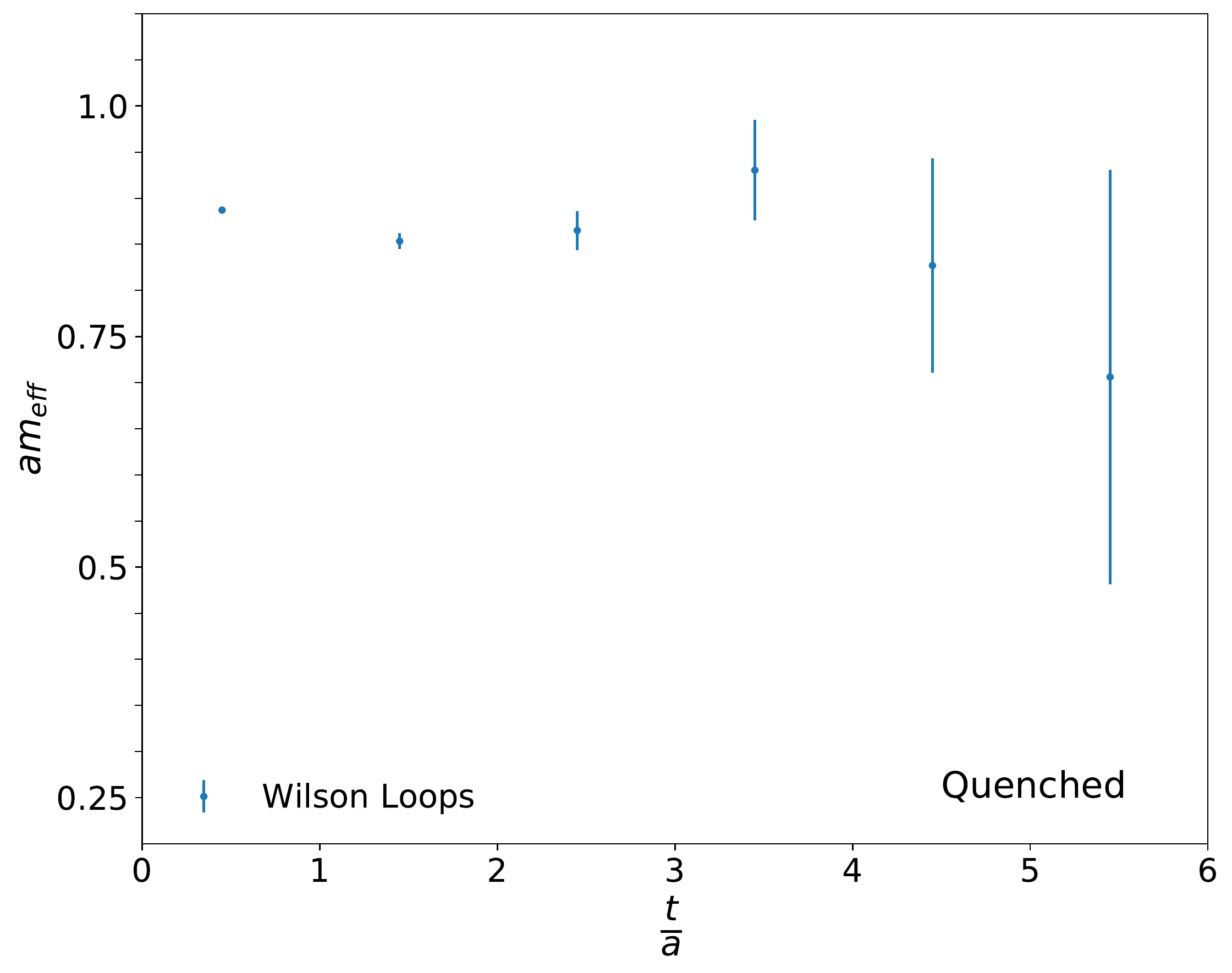}
\caption{Ground state effective mass of the scalar glueball made up of Wilson loops in the quenched setup.}
\label{Fig:QuenchedScalarGlueball}
\end{figure}

To complement the charmonia study done in this ensemble a preliminary study of the scalar glueball was performed. The glueball operators used in this work are built from Wilson loops whose shapes are chosen as to transform according to the irrep of interest and with the desired parity and charge conjugation symmetries, in this case $A_{1}^{++}$. The shapes used are those given in \cite{Berg1983} for length 8. To build a basis of operators for a GEVP formulation, single and double windings of these loops \cite{Morningstar1999}, together with those of the same shape but double length, were used and both HYP \cite{Hasenfratz2001} and improved APE \cite{Lucini2004} smearing schemes were tried. Using low statistics the optimal smearing parameters were determined and the most significant operators from the pruning were chosen as the variational basis. Fig. \ref{Fig:UnquenchedScalarGlueball} displays the resulting effective mass for the ground state of the scalar glueball in this work calculated using 10200 gauge configurations. A preliminary plateau can be seen yet there is still the presence of excited state contamination at very early times. To investigate the relation to the presence of the dynamical quarks the same measurements were performed in 7100 gauge configurations of a quenched ensemble with the same dimensions and almost equal $\frac{t_0}{a^2}$. This result can be seen in Fig. \ref{Fig:QuenchedScalarGlueball}, where there is a preliminary plateau with much less excited state contamination.

\section{Conclusions}
In this work an improvement to the distillation method was proposed via exploiting quark and meson distillation profiles. This improvement was tested by computing the spectrum of $N_f = 2$ QCD with mass at half the physical charm quark mass with both standard and improved distillation, displaying the clear advantages of the latter. Additionally the mass difference between the iso-scalar and iso-vector pseudoscalar operators was directly measured yielding a non-negligible result. To complement this study, the ground state mass of the scalar glueball was also measured in this setup resulting in a visible, albeit slightly contaminated by excited state contributions, signal.

\noindent \textbf{Acknowledgement.} The authors gratefully acknowledge the Gauss Centre for Supercomputing e.V. (www.gauss-centre.eu) for funding this project by providing computing time on the GCS Supercomputer SuperMUC-NG at Leibniz Supercomputing Centre (www.lrz.de).
\bibliographystyle{unsrtnat} 
\bibliography{refs}

\begin{thebibliography}{14}
\providecommand{\natexlab}[1]{#1}
\providecommand{\url}[1]{\texttt{#1}}
\expandafter\ifx\csname urlstyle\endcsname\relax
  \providecommand{\doi}[1]{doi: #1}\else
  \providecommand{\doi}{doi: \begingroup \urlstyle{rm}\Url}\fi

\bibitem[Peardon et~al.(2009)Peardon, Bulava, Foley, Morningstar, Dudek,
  Edwards, Jo{\'{o}}, Lin, Richards, and Juge]{Peardon2009}
Michael Peardon, John Bulava, Justin Foley, Colin Morningstar, Jozef Dudek,
  Robert~G. Edwards, B{\'{a}}lint Jo{\'{o}}, Huey-Wen Lin, David~G. Richards,
  and Keisuke~Jimmy Juge.
\newblock Novel quark-field creation operator construction for hadronic physics
  in lattice {QCD}.
\newblock \emph{Physical Review D}, 80\penalty0 (5), September 2009.
\newblock \doi{10.1103/physrevd.80.054506}.
\newblock URL \url{https://doi.org/10.1103/physrevd.80.054506}.

\bibitem[Morningstar et~al.(2011)Morningstar, Bulava, Foley, Juge, Lenkner,
  Peardon, and Wong]{Morningstar2011}
C.~Morningstar, J.~Bulava, J.~Foley, K.~J. Juge, D.~Lenkner, M.~Peardon, and
  C.~H. Wong.
\newblock Improved stochastic estimation of quark propagation with laplacian
  heaviside smearing in lattice {QCD}.
\newblock \emph{Physical Review D}, 83\penalty0 (11), June 2011.
\newblock \doi{10.1103/physrevd.83.114505}.
\newblock URL \url{https://doi.org/10.1103/physrevd.83.114505}.

\bibitem[Balog et~al.(1999)Balog, Niedermaier, Niedermayer, Patrascioiu,
  Seiler, and Weisz]{Balog1999}
J{\'{a}}nos Balog, Max Niedermaier, Ferenc Niedermayer, Adrian Patrascioiu,
  Erhard Seiler, and Peter Weisz.
\newblock Comparison of the o(3) bootstrap $\upsigma$ model with lattice
  regularization at low energies.
\newblock \emph{Physical Review D}, 60\penalty0 (9), October 1999.
\newblock \doi{10.1103/physrevd.60.094508}.
\newblock URL \url{https://doi.org/10.1103/physrevd.60.094508}.

\bibitem[Niedermayer et~al.(2001)Niedermayer, R\"{u}fenacht, and
  Wenger]{Niedermayer2001}
Ferenc Niedermayer, Philipp R\"{u}fenacht, and Urs Wenger.
\newblock Fixed point gauge actions with fat links: scaling and glueballs.
\newblock \emph{Nuclear Physics B}, 597\penalty0 (1-3):\penalty0 413--450,
  March 2001.
\newblock \doi{10.1016/s0550-3213(00)00731-8}.
\newblock URL \url{https://doi.org/10.1016/s0550-3213(00)00731-8}.

\bibitem[Michael and Teasdale(1983)]{Michael1983}
C.~Michael and I.~Teasdale.
\newblock Extracting glueball masses from lattice {QCD}.
\newblock \emph{Nuclear Physics B}, 215\penalty0 (3):\penalty0 433--446,
  February 1983.
\newblock \doi{10.1016/0550-3213(83)90674-0}.
\newblock URL \url{https://doi.org/10.1016/0550-3213(83)90674-0}.

\bibitem[L\"{u}scher and Wolff(1990)]{Lscher1990}
Martin L\"{u}scher and Ulli Wolff.
\newblock How to calculate the elastic scattering matrix in two-dimensional
  quantum field theories by numerical simulation.
\newblock \emph{Nuclear Physics B}, 339\penalty0 (1):\penalty0 222--252, July
  1990.
\newblock \doi{10.1016/0550-3213(90)90540-t}.
\newblock URL \url{https://doi.org/10.1016/0550-3213(90)90540-t}.

\bibitem[Blossier et~al.(2009)Blossier, Morte, von Hippel, Mendes, and
  Sommer]{collaboration2009}
Benoit Blossier, Michele~Della Morte, Georg von Hippel, Tereza Mendes, and
  Rainer Sommer.
\newblock On the generalized eigenvalue method for energies and matrix elements
  in lattice field theory.
\newblock \emph{Journal of High Energy Physics}, 2009\penalty0 (04):\penalty0
  094--094, April 2009.
\newblock \doi{10.1088/1126-6708/2009/04/094}.
\newblock URL \url{https://doi.org/10.1088/1126-6708/2009/04/094}.

\bibitem[L\"{u}scher(2010)]{Flow}
Martin L\"{u}scher.
\newblock Properties and uses of the wilson flow in lattice {QCD}.
\newblock 2010\penalty0 (8), August 2010.
\newblock \doi{10.1007/jhep08(2010)071}.
\newblock URL \url{https://doi.org/10.1007/jhep08(2010)071}.

\bibitem[Dudek et~al.(2008)Dudek, Edwards, Mathur, and Richards]{Dudek2008}
Jozef~J. Dudek, Robert~G. Edwards, Nilmani Mathur, and David~G. Richards.
\newblock Charmonium excited state spectrum in lattice {QCD}.
\newblock \emph{Physical Review D}, 77\penalty0 (3), February 2008.
\newblock \doi{10.1103/physrevd.77.034501}.
\newblock URL \url{https://doi.org/10.1103/physrevd.77.034501}.

\bibitem[Hatton et~al.(2020)Hatton, Davies, Galloway, Koponen, Lepage, and
  and]{Hatton2020}
D.~Hatton, C.{\hspace{0.167em}}T.{\hspace{0.167em}}H. Davies, B.~Galloway,
  J.~Koponen, G.{\hspace{0.167em}}P. Lepage, and A.{\hspace{0.167em}}T.~Lytle
  and.
\newblock Charmonium properties from lattice {QCD}+{QED} : Hyperfine splitting,
  j/$\uppsi$ leptonic width, charm quark mass, and $a_{\upmu}^c$.
\newblock \emph{Physical Review D}, 102\penalty0 (5), September 2020.
\newblock \doi{10.1103/physrevd.102.054511}.
\newblock URL \url{https://doi.org/10.1103/physrevd.102.054511}.

\bibitem[Berg and Billoire(1983)]{Berg1983}
B.~Berg and A.~Billoire.
\newblock Glueball spectroscopy in 4d {SU}(3) lattice gauge theory (i).
\newblock \emph{Nuclear Physics B}, 221\penalty0 (1):\penalty0 109--140, July
  1983.
\newblock \doi{10.1016/0550-3213(83)90620-x}.
\newblock URL \url{https://doi.org/10.1016/0550-3213(83)90620-x}.

\bibitem[Morningstar and Peardon(1999)]{Morningstar1999}
Colin~J. Morningstar and Mike Peardon.
\newblock Glueball spectrum from an anisotropic lattice study.
\newblock \emph{Physical Review D}, 60\penalty0 (3), July 1999.
\newblock \doi{10.1103/physrevd.60.034509}.
\newblock URL \url{https://doi.org/10.1103/physrevd.60.034509}.

\bibitem[Hasenfratz and Knechtli(2001)]{Hasenfratz2001}
Anna Hasenfratz and Francesco Knechtli.
\newblock Flavor symmetry and the static potential with hypercubic blocking.
\newblock \emph{Physical Review D}, 64\penalty0 (3), July 2001.
\newblock \doi{10.1103/physrevd.64.034504}.
\newblock URL \url{https://doi.org/10.1103/physrevd.64.034504}.

\bibitem[Lucini et~al.(2004)Lucini, Teper, and Wenger]{Lucini2004}
Biagio Lucini, Michael Teper, and Urs Wenger.
\newblock Glueballs andk-strings in {SU}(n) gauge theories: calculations with
  improved operators.
\newblock \emph{Journal of High Energy Physics}, 2004\penalty0 (06):\penalty0
  012--012, June 2004.
\newblock \doi{10.1088/1126-6708/2004/06/012}.
\newblock URL \url{https://doi.org/10.1088/1126-6708/2004/06/012}.

\end{thebibliography}

\end{document}